\documentstyle[12pt]{article}
\newcommand{\be}{\begin{equation}}
\newcommand{\ee}{\end{equation}}
\newcommand{\ba}{\begin{eqnarray}}
\newcommand{\ea}{\end{eqnarray}}
\def\n{\noindent}

\catcode `\@=11
\@addtoreset{equation}{section}
 
\makeatletter
\@addtoreset{equation}{section}
\renewcommand\theequation
{\ifnum \c@section>\z@ \thesection .\fi \@arabic\c@equation}
\makeatother

\catcode `\@=12

\title{\bf\huge Domain Walls in Kaluza-Klein spacetime                                     }     
\author{L.K. Patel \\
{\sl Department of Mathematics,}\\
{\sl Gujarat University, Ahmedabad 380 009. India} \\
Naresh Dadhich\thanks{E-mail : nkd@iucaa.ernet.in} \\
{\sl Inter-University Centre for Astronomy \& Astrophysics,}\\
{\sl Post Bag 4, Ganeshkhind, Pune - 411 007, India.} \\
R. Tikekar \\
{\sl Department of Mathematics,}\\
{\sl Sardar Patel University, Vallabh Vidyanagar 388 120, India.} \\
}
\date{}
\begin{document}
\maketitle

\begin{abstract}
Three families of exact solutions of Einstein field equations are
found. Each family contains three parameters. Two of these
families represent thick domain walls in a five dimensional
Kaluza-Klein spacetime. The dynamical behaviour of our models
is briefly discussed. The spacetime in all the cases is found
to be reflection symmetric with respect to the wall.
\end{abstract}
\newpage
\section{Introduction}

The phase transitions in the early universe, due to spontaneous 
breaking of a discrete symmetry, could have produced the topological
defects such as domain walls, strings and monopoles [1].
Hill, Schramm and Fry [2] have suggested that light domain walls
of large thickness may have been produced during the late time phase
transitions such as those occuring after the decoupling of matter
and radiation. Recently the study of the thick domain walls and
spacetimes associated with them have received considerable attention
due to their application in structure formation in the Universe.

\n Vilenkin [3] first showed that the gravitational field fo an infinite
thin domain wall with planar symmetry cannot be described by a
static metric. Subsequently, Widrow [4] noted that nor could  a thick domain 
wall be described by a regular static metric.
These considerations suggest that non-static metrics are suitable
for description of the field of a thick domain wall. Many
authors have discussed non-static solutions of the Einstein scalar
field equations for thick domain walls [4-6]. But these solutions
have peculiar behaviour. In these solutions the energy scalar is independent
of time whereas the metric tensor depends on both space and time. 
Letelier and Wang [7] have obtained exact solutions to the Einstein field
equations that represent the collision of plane thin walls.

\n Later on Wang [8] has derived a two parameter family of solutions
of the Einstein field equations representing gravitational collapse of a 
thick domain wall. Thick domain walls are characterized by the
energy momentum tensor

\be
T_{ik} = \rho (g_{ik} + w_i w_k) + p w_i w_k, ~w_i w^i = -1
\ee

\n where $\rho$ is the energy density of the wall, $p$ is the pressure
in the direction normal to the plane of the wall and $w_i$ is a unit
spacelike vector in the same direction. In Wang's solution, the energy
scalar and the metric tensor are dependent on space as well as time
coordinates.

\n There are two approaches for the study of thick domain walls. In the
first approach one studies the field equations as well as the equations
of domain wall treated as the self-interacting scalar field. In the
second approach one assumes the energy momentum tensor in the form
(1.1) and then the field equations are solved. The second approach seems
to be easier. In the present paper we adopt the second approach
but apply it to a more general five dimensional Kaluza-Klein spacetime.
The advances in supergravity in 11-D and superstring in 10-D indicate
that the multidimensionality of space is apparently a fairly adequate
reflection of the dynamics of interactions over the distance where 
all forces unify.
The cosmological implications of higher dimensions were first discussed
by Chodos and Detweiler [9]. They have Kasner-type vaccum solutions
in a five dimensional spacetime. Their solutions possess the property
of dimensional reduction. It would be worthwhile to study the time-dependent
thick domain walls in a five dimensional Kaluza-Klein spacetime. Banerjee
and Das [10] have considered thick domain walls in higher dimensions
and obtained some exact solutions. The purpose
of the present work is to report some other new exact solutions
of the Einstein equations representing gravitation field of thick domain
walls in a five-dimensional spacetime.

\n In Sec.2 we set up the field equations and solve them and we conclude 
in Sec.3 with a discussion. 

\section{Field equations and their solutions}

In this section, we engage ourselves to the problem of construction
of general relativistic models of plane symmetric thick domain
walls in a five-dimensional spacetime. Here the spacetime admits one
additional killing vector. The general five-dimensional plane
symmetric metric can be expressed in the form

\be
ds^2 = A^2 (dt^2 - dx^2) - B^2(dy^2 + dz^2) - E^2 d \psi^2
\ee

\n where $A,B$ and $E$ are functions of $x$ and $t$, and $\psi$
is the fifth coordinate corresponding to the extra dimension.

We introduce the pentad

\be
\theta^0 = Adt, \theta^1 = Adx, \theta^2 = Bdy, \theta^3 = Bdz,
\theta^4 = E d \psi.
\ee

\n Here and in what follows all the components would refer to base frame.
The surviving Ricci components $R_{ab}$ are listed below for ready reference:

\ba
A^2 R_{00} &=& \frac{\ddot A}{A} + \frac{2 \ddot B}{B} +
\frac{\ddot E}{E} - \frac{\dot A}{A} (\frac{\dot A}{A}
+ \frac{2 \dot B}{B} + \frac{\dot E}{E}) - \nonumber \\
&&[\frac{A^{\prime\prime}}{A} + \frac{A^{\prime}}{A} (\frac{2B^{\prime}}{B} +
\frac{E^{\prime}}{E} - \frac{A^{\prime}}{A})]
\ea

\be
A^2 R_{01} = \frac{2 \dot B^{\prime}}{B} + \frac{\dot E^{\prime}}{E} -
\frac{\dot A}{A}(\frac{2 B^{\prime}}{B} + \frac{E^{\prime}}{E})
- \frac{A^{\prime}}{A}(\frac{2 \dot B}{B} + \frac{\dot E}{E})
\ee

\ba
A^2 R_{(11)} &=& \frac{A^{\prime \prime}}{A} + \frac{2 B^{\prime \prime}}{B}
+ \frac{E^{\prime \prime}}{E} - \frac{A^{\prime}}{A} (\frac{A^{\prime}}{A}
+ \frac{2 B^{\prime}}{B} + \frac{E^{\prime}}{E}) \nonumber \\
&&- [\frac{\ddot A}{A}
+ \frac{\dot A}{A} (2 \frac{\dot B}{B} + \frac{\dot E}{E} - \frac{\dot A}{A})]
\ea

\be 
A^2 R_{22} = A^2 R_{33} = \frac{B^{\prime \prime}}{B} +
\frac{B^{\prime}}{B} (\frac{B^{\prime}}{B} + \frac{E^{\prime}}{E})
- [\frac{\ddot B}{B} + \frac{\dot B}{B} (\frac{\dot B}{B} +
\frac{\dot E}{E})]
\ee

\be
A^2 R_{44} = \frac{E^{\prime \prime}}{E} + 2 \frac{E^{\prime}B^{\prime}}
{EB} - (\frac{\ddot E}{E} + \frac{2 \dot B \dot E}{BE})
\ee

\n where a prime and a dot indicate derivatives with respect to $x$ and $t$
respectively.

\n We now have to solve the Einstein field equations

\be
R_{ab} = - 8 \pi [T_{ab} - \frac{1}{3} Tg_{ab}]
\ee

\n where the energy stress components in the comoving coordinates for the 
thick domain wall are given by

\be
T^0_0 = T^2_2 = T^3_3 = \rho, T^1_1 = -p, T^0_1 = 0, T^4_4 = 0
\ee

\n Here $\rho$ is the energy density of the wall which is also equal to the
tension along $y$ and $z$ directions in the plane of the wall, $p$ is the
pressure along x-direction. The stress component $T^4_4$ corresponding to
the extra dimension is assumed to be zero. In view of (2.9), equations
(2.8) lead ot the following relations: 

\be 
R_{01} = 0
\ee

\be
R_{22} = -R_{00}
\ee

\be
3R_{22} + R_{11} - R_{44} = 0
\ee

\be
8 \pi p = 3R_{22}
\ee

\be
8 \pi \rho = -(R_{11} + 2 R_{22}).
\ee

\n The general solution of the above system of equations is quite 
difficult to obtain. So we make the following separability assumptions 
for the metric potentials:

\be
A = cosh^a (mx) e^{\alpha kt}, B = cosh^b (mx) e^{\beta kt},
E = cosh^d (mx) e^{\delta kt}.
\ee

\n Here $a, b, d, \alpha, \beta, \delta, m$ and $k$ are real constants.
With these assumptions, equation (2.10) leads to

\be
2 \beta (b-a) - \alpha (d + 2b) + \delta(d-a) = 0.
\ee

\n $R_{(22)} = -R_{(00)}$ gives

\be
(a-b) m^2 [(2b +d-1) sech^2 (mx) - (2b+d)] + k^2
[\delta^2 - \delta (\alpha + \beta) - 2 \alpha \beta] = 0.
\ee

\n Eqn. (2.12) on simplification leads to

\be
m^2 [8b^2 - 2ab -ad +bd]
+ m^2 sech^2 (mx) [a + 5b - 8b^2 + 2ab + ad - bd]
= k^2 (2 \beta + \delta) (\alpha + 3 \beta - \delta).
\ee

\n Eqns.. (2-13) and (2-14) determine the physical
parameters $p$ and $\rho$. They are given by

\be
\frac{8 \pi p A^2}{3} = bm^2 [(2b+d) + (1-2b-d)
sech^2 (mx)] - \beta (2 \beta + \delta) k^2
\ee

\n and

\ba
- 8 \pi \rho A^2 &=& m^2 (6 b^2 + d^2 - 2ab + bd - ad) \nonumber \\
&+& m^2 sech^2 (mx) [- 6 b^2 - d^2 + 2ab + ad - 2bd
+ a + d + 4b] \nonumber \\
&-& (\alpha + 2 \beta)(2 \beta + \delta) k^2.
\ea

\n From eqn. (2.17) it is clear that $(a-b)(2b+d-1) = 0$, which will ipmply
either $a=b$ or $2b +d =1$, and correspondingly we will have the 
following two cases.

\n {\bf Case I:} $a=b$.

\n In this case, eqns (2.16) - (2.18) give

\be
a = b = 1, ~\frac{m^2}{k^2} = \beta^2, 
~\alpha = \frac{d(d-1)}{(d+2)} \beta, ~\delta = \beta d 
\ee

\n where $d, \beta, $ and $k$ are arbitrary parameters. The
pressure $p$ and the energy density $\rho$ are given by

\be
8 \pi p = -3 m^2 (d+1) sech^4 (mx) e^{-2 \alpha kt}
\ee

\n and

\be
8 \pi \rho = m^2 (d^2 -2) sech^4 (mx) e^{-2 \alpha kt}.
\ee

\n Therefore

\be
\rho = - \frac{(d^2 - 2)}{3 (d+1)} p.
\ee

\n Thus we have a three parameter family of solutions describing thick
domain walls. When $\beta = -1/2$, it reduces
to the two parameter family of solutions discussed by Banerjee and Das [10]
with slight change of notations.

\n{\bf Case II:} $2b+d=1$.

\n Eqns (2.16) - (2.18) give

\ba
a = b(3b -2), ~d = 1-2b, \nonumber \\
\alpha = 6b(1-b) \beta + (1-3b^2) \delta, \nonumber \\
-3b(1-b) m^2 = k^2 (\delta^2 - 2 \alpha \beta - \alpha \delta
- \beta \delta), \nonumber \\
-3b (1+b) m^2 = k^2 (\delta - \alpha - 3 \beta) (2 \beta + \delta).
\ea

\n The last two equations lead  to 

\be
(b-1) (1-4b^2) \frac{\beta^2}{\delta^2} + b(1 + 2b - 4b^2) \frac{\beta}
{\delta} - b^3 = 0
\ee

\n which has the following two roots

\be
\frac{\beta}{\delta} = \frac{b}{1-2b}, ~\frac{b^2}{(1-b) (2b+1)}.
\ee

\n Therefore we have to consider two separate cases.

\n {\bf Case II (i)}

\n In this case we obtain

\ba
\beta &=& \frac{b \delta}{1-2b}, ~\alpha = \frac{(1-2b + 3b^2) \delta}
{(1-2b)}, ~d = 1-2b \nonumber \\
a &=& b(3b-2), ~\frac{m^2}{k^2} = \frac{(1-b + 4b^2)}{(1+b)(1-2b)} \delta^2
\ea

\n and

\be
\frac{8 \pi p}{3} = \frac{2k^2 b^2 \delta^2 (3b - 4b^2 -2)}{(1-2b)^2 (1+b)}
sech^{2a} (mx) e^{-2 k \alpha t}
\ee

\n and

\be
8 \pi \rho =  \frac{b\delta^2k^2(1+3b^2)(8b^2 - 12b + 7)}{(1-2b)^2 
(1+b)}sech^{2a} (mx) e^{-2 k \alpha t}
\ee

\n Here also the ratio $\rho/p$ is a constant.

\n {\bf Case II (ii)}

\n For this case we get

\ba
d&=&1-2b, ~a=b(3b-2), ~\beta= -\frac{b^2 \delta}{(b-1)(2b+1)} \nonumber \\
\alpha&=&\frac{(1-b)(1+3b)}{(1+2b)}, ~\frac{m^2}{k^2} =
\frac{b(2b^2 - b - 2)}{(b-1)^2(2b+1)^2} \delta^2.
\ea

\n The pressure and density are given by

\be
8 \pi p = \frac{3b^2 k^2 \delta^2 (2b^2 - 2b -3)}{(b-1)^2(2b+1)^2}
sech^{2a} (mx) e^{- 2 \alpha kt}
\ee

\n and 

\be
8 \pi \rho = \frac{k^2 \delta^2 (-6b^5 + 6b^4 + 4b^3 - b^2 + 4b + 1)}  
{(b-1)^2(2b+1)^2} sech^{2a} (mx) e^{- 2 \alpha kt} 
\ee

\n and they again bear constant ratio.

\n In case II, we thus obtain two three parameter families of solutions for 
domain walls, the parameters being $b, \delta$ and $k$.

\section{Discussion}

It is clear that the spacetimes of solutions of Case I and Case II
are reflection symmetric with respect to the wall. For a thick domain wall
it is desirable that pressure and density decrease on both sides of the wall 
away from the symmetry plane and fall off to zero as $x \rightarrow 
\pm \infty$.

\n For the domain wall solution of Case I the physical requirements $\rho > 0,
p > 0$ and $\rho - p \geq 0$ would be satisfied provided we choose the 
parameter
$d$ such that $d < -(3 + \sqrt{5})/2$. When $d = -(3 + \sqrt{5})/2$ then 
it is $p = \rho$. Clearly $\rho,p$ fall off to zero on either side of 
the wall.

For the solutions of Case II(i) and Case II(ii), the proper fall off 
behaviour would require $a>0$. In Case II(i), this requirement would 
conflict with $\rho>0, m^2/k^2\ge0$. Thus this family is not physically 
viable. However 
it is interesting to note that when $b=0$, $\rho$ and $p$ vanish,  
resulting into an empty spacetime given by the metric,

\be
ds^2 = e^{2nt} (dt^2 - dx^2) - dy^2 - dz^2 - e^{2nt} cosh^2 (nx)d \psi^2
\ee

\n where we have set $n= k\delta$.

\n It can be easily checked that when $b< \frac{1}{4} (1- \sqrt{17}) $
(i.e. $2b^2 - b - 2 \leq 0)$, then the solutions of Case II (ii) would 
satisfy
the requirements $a>0, \rho > 0$ and $m^2/k^2 > 0$. It would have 
the proper fall off behaviour as well as $\rho - p \geq 0$.

\n We shall now discuss the dynamical behaviour of our models under 
different restrictions imposed on various parameters occuring in the 
solutions. The general expression for the three space volume is given by

\be
|g_{3}|^{1/2} = cosh^{a+2b} (mx) e^{kt(\alpha + 2 \beta)}
\ee

\n Thus for temporal behaviour would be

\be
|g_{3}|^{1/2} \sim exp [kt ( \alpha + 2 \beta)]
\ee

\n Here it should be noted that when $\beta = -1/2$ in C
ase I, 
we recover the Banerjee and Das solution [10].
So, for case I, we take $\beta$ to be negative. If $d < -(3 + 
\sqrt{5})/2$, then we have $\alpha + 2 \beta < 0$. Further if $k>0$, 
3-space  collapses while the extra dimension inflates. In this 
process
we get a singularity because as $t \rightarrow \infty$, $\rho$ and $p$
diverge. On the other hand if $k < 0$, the effective  3-space inflates 
while the extra dimension collapses in course of time.

\n On the similar lines we can discuss the dynamical behaviour of the domain 
wall solutions of Case II (ii). For the sake of brevity we shall not
go into these details here.

\n The repulsive and attractive character of thick domain walls can be 
discussed by either studying the timelike geodesics in the spacetime or 
analysing the acceleration of an observer who is at rest relative to the 
wall [11]. Let us consider an observer with the 
four velocity $v_i = cosh^a (mx) e^{\alpha kt} \delta^t_i$. Then we 
obtain the acceleration vector $A^i$ as

\be
A^i = v^i_{;k} v^k = am tanh (mx) cosh^{-2a} (mx) e^{-2 \alpha kt}
\delta^i_x.
\ee

\n In case I, $a=1$, and  if $m>0$, then $A^x$ is 
positive. This implies that in order to keep the observer comoving
with the wall it has to accelerate away from the symmetry plane or 
in other words it is attracted towards the wall. Similarly if $m < 0$, then 
the wall exhibites a repulsive nature to the 
observer. Similar conclusions can be drawn for the domain wall solutions
of case II (ii).

\n If we assume

\be
T^0_0 = T^2_2 = T^3_3 = T^4_4 = \rho, T^1_1 = -p, T^0_1 = 0
\ee

\n instead of (2.9), we get a domain wall solution in which $\rho=p$.
But this solution is the same as that given by Banerjee and Das [10].

\n The function $e^{x^2}$ is also reflection symmetric about the
yz-plane. If we take $e^{mx^2}$ in place of $cosh (mx)$ in the 
separability assumption (2.15), there cannot occur any domain wall 
solutions. But it does give a five-dimensional empty spacetime described by 
the metric

\be
ds^2 = e^{2nt + n^2 x^2} (dt^2 - dx^2) - e^{\frac{2nt}{\sqrt{6}}}
(dy^2 + dz^2) - e^{\frac{4nt}{\sqrt{6}}} d \psi^2
\ee

\n where $n$ is an arbitrary constant, and is the cause for the 
spacetime curvature. It is an inhomogeneos vacuum spacetime.  \\

\n {\bf Acknowledgements:}

LKP and RT thank IUCAA, Pune for hospitality during their visits there.


\newpage


\begin{thebibliography}{99}
\bibitem{}A. Vilenkin, Phys. Rep. {\bf 121}, 263 (1985).
\bibitem{} C.T. Hill, D.N. Schramm and J.N. Fry, Nucl. Part. Phys.,
{\bf 19}, 25 (1989).

\bibitem{} A. Vilenkin, Phys. Rev. {\bf D23}, 852 (1981).

\bibitem{} L.M. Widrow, Phys. Rev. {\bf D 39}, 3571 (1989).

\bibitem{} G. Goetz, J. Math. Phys. {\bf 31}, 2683 (1990).

\bibitem{} M. Mukherjee, Class. Quantum Grav., {\bf 10}, 131 (1993).

\bibitem{} P.S. Letelier and A. Wang, Class. Quantum Grav. {\bf 10},
L-29 (1993).

\bibitem{} A. Wang, Mod. Phys. Lett. {\bf A90}, 3605 (1994).

\bibitem{} A. Chodos and S. Detweider, Phys. Rev. {\bf D21}, 2167
(1980).

\bibitem{} A. Banerjee and A. Das, Int. J. Mod. Phys. {\bf D7}, 81 (1998).

\bibitem{} A. Wang, Phys. Rev. {\bf D 45}, 3534 (1992).
\end{thebibliography}
\end{document}